\documentclass[showpacs,showkeys,preprintnumbers,amsmath,amssymb,twocolumn]{revtex4-1}

\usepackage{graphicx}
\usepackage{color}

\begin{document}

\title{Electronic transport properties of (fluorinated) metal phthalocyanine}
\author{M.\ M.\ Fadlallah$^{1,2,3}$}\
\author{U. Eckern$^1$}\
\author{A.\ H.\ Romero$^4$}\
\author{U.\ Schwingenschl\"ogl$^5$}\
\affiliation{
$^1$Institut f\"ur Physik, Universit\"at Augsburg, 86135 Augsburg, Germany\\
$^2$Physics Department, Faculty of Science, Benha University, Benha, Egypt\\
$^3$Centre for Fundamental Physics, Zewail City of Science and Technology, Giza, Egypt\\
$^4$Physics Department, West Virginia University, 26506-6315, Morgantown, USA\\
$^5$KAUST, PSE Division, Thuwal 23955-6900, Kingdom of Saudi Arabia}

\begin{abstract}
The magnetic and transport properties of the metal phthalocyanine (MPc) and F$_{16}$MPc 
(M = Sc, Ti, V, Cr, Mn, Fe, Co, Ni, Cu, Zn and Ag) families of molecules in contact with S-Au wires are investigated 
by density functional theory within the local density approximation, including local electronic correlations on the central
metal atom. The magnetic moments are found to be considerably modified under fluorination. In addition, they do 
not depend exclusively on the configuration of the outer electronic shell of the central metal atom (as in isolated
MPc and F$_{16}$MPc) but also on the interaction with the leads.
Good agreement between the calculated conductance and experimental results is obtained.
For M = Ag, a high spin filter efficiency and conductance is observed, giving rise to a potentially high sensitivity
for chemical sensor applications.
\end{abstract}


\maketitle

\section{Introduction}

Metal phthalocyanines (MPcs) constitute a family of medium sized molecular semiconductors, which are of considerable interest for
numerous applications such as chemical sensors \cite{Wright}, fuel cells \cite{Baranton}, solar cells \cite{Seguy},
and optoelectronic devices \cite{Hung}. A large number 
of MPcs can be synthesized \cite{Ali} and geometrically arranged in large areas at low cost \cite{book1}, which is important from the 
technological point of view. In particular, the photoconductivity of MPcs has been studied intensively 
with the purpose of increasing the electrical conductivity of devices based on these molecules \cite{A1A}. The improvement of the 
organic semiconductor devices relies on the quality of the metal-organic interfaces \cite{Peisert}, in particular, on 
the efficiency of charge injection and on the mobility of the charge carriers \cite{Ozaki,Song}. For example, there is a charge 
transfer shift of the electronic levels at the CuPc/Au interface in the early stages of CuPc deposition on Au \cite{PeisertA}, and
new occupied molecular states are created \cite{Schwieger,Schwieger1}.

The replacement of the M atom as well as the substitution of H by F (fluorinated MPc) alters the gap between
the highest occupied molecular orbital (HOMO) and the lowest unoccupied molecular orbital \cite{Leznoff},
the latter resulting in n-type conduction, in contrast to the p-type behavior of the standard MPcs \cite{Huang,TWang}. The charge transport
through the molecule can be measured directly \cite{Kan,Kig} or indirectly by photo-induced electron 
transfer \cite{Gul}, time resolved microwave conductivity \cite{Win}, and scanning tunneling microscopy \cite{Sed}.
CuPc has been contacted with Au atomic chains on NiAl substrate by Nazin et al.\ \cite{Naz}, who observed a shift and splitting of the molecular orbitals 
as well as modifications of the electrode orbitals by scanning tunneling microscopy.
Field-effect transistors and metal-insulator-semiconductor diodes have been used to study the transport
through CuPc for different leads such as Ca, Au, and F$_{4}$TCNQ/Au, demonstrating both electron or hole transport with a 
strong dependence on the geometry of the molecule-metal contact \cite{Co}. CuPc sandwiched between two semi-infinite Au electrodes 
has been investigated theoretically in Refs.\ \cite{Tada2,Tada1}. The transmission coefficient, $T(E)$, shows two peaks near the Fermi energy 
($E_F$) which have been dissected in terms of molecular orbitals. The electronic states of CuPc hardly change when leads are attached.

On the other hand, the Landauer approach and Green's function formalism have been used to address the quantum transport in MPc 
structures. MPcs with M $=$ Mn, Fe, Co, Ni, Cu and Zn sandwiched
between semi-infinite armchair single-wall carbon nanotubes (SWNT) have been considered in Ref.\ \cite{Shen} using the SMEAGOL 
package \cite{Rocha,Rungger}. Within the generalized gradient approximation (GGA) it has been 
concluded that FePc and MnPc can be used as spin filters, and that the spin filter efficiency increases when N-doped graphene 
nanoribbons are used as leads for FePc \cite{Huan}. To overcome the weak interaction between Au contacts and MPcs, S atoms have been 
added, leading to a distinct molecular bonding. The transport properties of such AuS-MPc-SAu (M $=$ Cu, Mn) systems have 
been studied by the WanT code, with the result that electronic correlations are likely to be irrelevant \cite{Calz}.
Additionally, the authors have concluded that CuPc is a molecular conductor, and MnPc a spin filter.

In this article we extend previous works on AuS-MPc-SAu and AuS-F$_{16}$MPc-SAu junctions by including
spin polarization, and by considering a variety of different metals (M $=$ Sc, Ti, V, Cr, Mn, Fe, Co, Ni, Cu, Zn and Ag).
The effect of local electronic interactions on the central metal atom are also studied systematically.
In Sec.\ II we discuss the computational method, and present in Sec.\ III the magnetic properties. Then (Sec.\ IV) the
transmission properties and finally (Sec.\ V) the spin filter efficiency and electronic conductance are discussed. A summary
is given in the concluding Sec.\ VI.

\section{Computational method}

The transport properties are investigated using non-equilibrium Green's function
and density functional theory as implemented in the SMEAGOL \cite{Rocha,Rungger} and
SIESTA packages \cite{Sol}. The wave functions are expanded in atomic orbitals with
an energy cutoff of 300 Ry \cite{note}. We use a double zeta plus polarization basis.
Test calculations show no significant change when a single zeta basis is used for the
M atom; however, when a single zeta basis set is used for the whole structure, the results
are clearly of inferior quality. The nuclei 
and core electrons are represented by Troullier-Martins pseudopotentials \cite{Troullier}.
We employ the local density approximation with Coulomb interaction (LDA+$U$). The on-site Coulomb interaction, $U$, is applied to 
the d orbitals on the central metal ions. Forces below 0.04 eV/\AA\ are achieved in the structural optimization of MPc, using the conjugate 
gradient method. Since for metallic chains a Peierls gap is expected around the Fermi energy, we do not relax the structure 
after attaching the leads to the molecule \cite{Inder}.
We use the first two of the five-Au-atom leads, respectively, to calculate the surface Green's function, while the other
three, closer to the molecule,
are assumed to be part of what is usually called ``scattering region''. The Au-Au distance is chosen to be 
2.89 \AA\ \cite{Co,Calz,Naz}, and the interface distance between MPc and Au is set to 1.37 \AA\  \cite{Co}.

\begin{figure}[htb]
\includegraphics[width=0.48\textwidth]{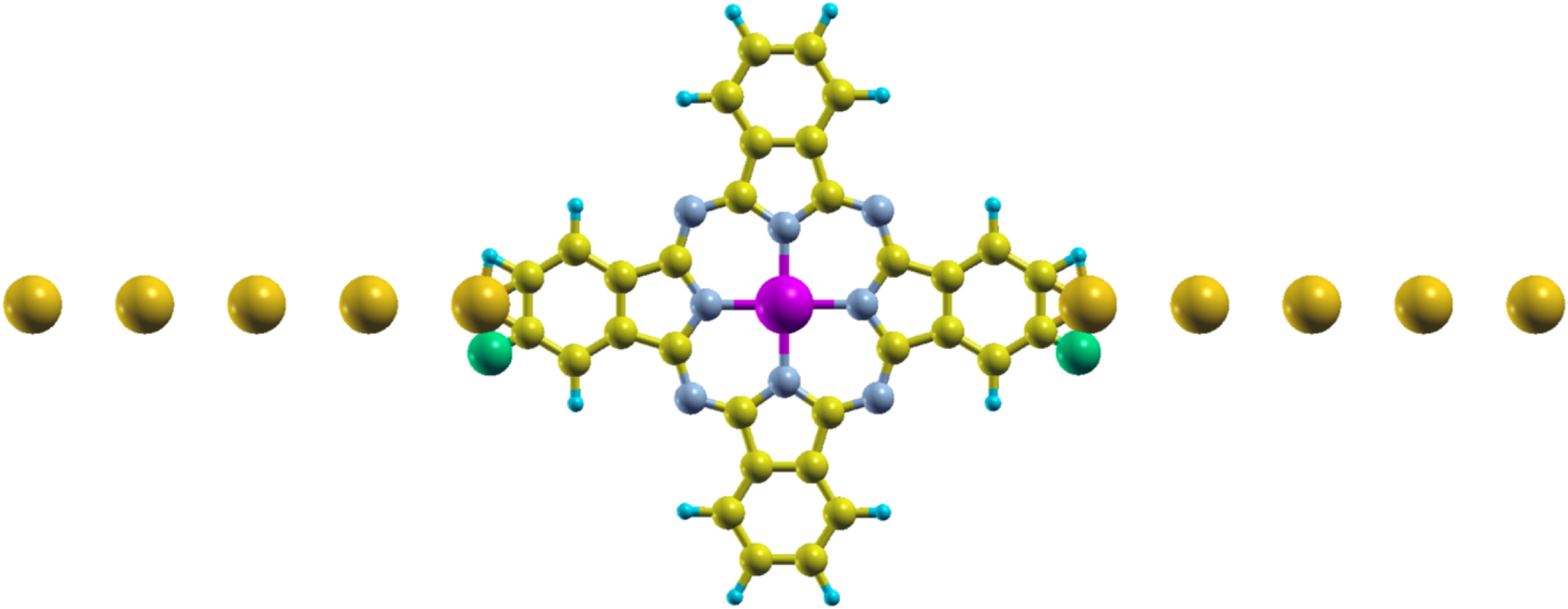}
\caption{(Color online) Structure of AuS-MPc-SAu.
H, C, N, S, M and Au atoms are shown in cyan, yellow, grey, green, pink and golden, respectively.}
\label{fig1}
\end{figure}

The structure of the devices is shown in Fig.\ \ref{fig1}. As a general remark, we note that the Mulliken analysis 
shows a considerable charge redistribution due to the leads (Au atoms), mainly 
involving the M atom, and the N and C atoms close to the central atom and at the contact between leads and molecule.

\section{Magnetic moments}

The electronic and magnetic properties of {\em isolated} MPc and F$_{16}$MPc molecules have been studied 
before \cite{Arillo}, where it was found that the magnetic moment (MM) is carried mainly by the metal atom. Below
we will compare, in particular, isolated molecules with molecules connected to leads, with emphasis on the
differences. In addition, since the central atom is a transition metal, we will encounter
all possible d orbital occupations, which often give rise to considerable correlation effects. Thus it
will be important to also investigate, not only for a few selected cases but systematically, the effect of
a local Coulomb interaction on the electronic structure as well as on transport properties.
In this context, we note that $U$ often is considered to be a parameter, to be determined by
comparing specific physical quantities, such as the band gap or the magnetic moment, with experiment.
However, the value obtained for $U$ in such a way depends on the quantity considered. On the other hand,
$U$ can in principle be computed by constrained density functional theory. Both approaches appear not
to be useful for the present systematic study where the system properties are strongly modified by the
coupling to the leads. Thus our concept is to vary $U$ systematically, within a reasonable range (from
$U=0$ to 8 eV, i.e., from weak to strong correlation) in order to elucidate the interaction-dependent
trends in the changes of the system properties. For this goal, it appears sufficient to rely on the three
values 0, 4, and 8 eV.

\begin{table*}[t]
\caption{Total and metal magnetic moments (in $\mu_{B}$) of 
AuS-MPc-SAu and AuS-F$_{16}$MPc-SAu at different $U$}
\centering
{
\begin{tabular}{|c|c|c|c|c|c|c|c|c|c|c|c|c|c|}\hline

& & & Sc & Ti & V & Cr & Mn & Fe & Co & Ni & Cu & Zn & Ag \\\hline

&$U=0$& total & 0.8 & 2.9 & 2.1 & 2.3 & 5.3 & 5.9 & 1.6 & 1.5 & 2.2 & 0.4 & 1.6 \\
& & metal &0.1 & 0.8 & 2.9 & 4.0 & 4.6 & 4.1 & 1.3 & 0.1 & 0.5 & 0.0 & 0.2 \\

MPc&$U=4$& total & 0.8 & 3.9 & 3.2 & 3.7 & 5.3 & 5.9 & 4.6 & 3.2 & 1.8 & 0.4 & 1.9 \\
& & metal &0.0 & 1.1 & 3.0 & 4.2 & 4.8 & 4.2 & 2.7 & 1.6 & 0.5 & 0.0 & 0.2 \\

&$U=8$& total & 0.8 & 3.8 & 3.2 & 4.1 & 5.3 & 5.8 & 4.4 & 2.9 & 1.6 & 0.4 & 2.1 \\
& & metal &0.0 & 1.1 & 3.0 & 4.2 & 4.9 & 4.6 & 2.8 & 1.7 & 0.5 & 0.0 & 0.2 \\ \hline

&$U=0$& total & 0.8 & 1.6 & 1.2 & 4.0 & 4.7 & 4.1 & 1.0 & 0.0 & 1.1 & 0.0 & 0.7 \\
& & metal &0.0 & 1.5 & 2.4 & 4.0 & 4.7 & 4.1 & 1.1 & 0.0 & 0.5 & 0.0 & 0.2  \\

F$_{16}$MPc&$U=4$& total & 0.8 & 1.5 & 3.2 & 4.0 & 4.9 & 3.4 & 2.9 & 1.9 & 1.1 & 0.0 & 0.8 \\
& & metal &0.0 & 1.5 & 2.6 & 4.1 & 5.0 & 4.2 & 2.6 & 1.5 & 0.5 & 0.0 & 0.3 \\

&$U=8$& total & 0.8 & 1.4 & 3.3 & 4.0 & 5.0 & 3.4 & 3.0 & 1.9 & 1.1 & 0.0 & 0.9 \\
& & metal &0.0 & 1.3 & 2.3 & 4.1 & 5.0 & 4.3 & 2.8 & 1.6 & 0.5 & 0.0 & 0.3 \\\hline

\end{tabular}
}
\label{table1}
\end{table*}

The calculated magnetic moments are summarized in Table \ref{table1}. We first discuss the total MMs for the 
AuS-MPc-SAu system. For M = Sc the MM differs slightly from that of the isolated molecule ($1\,\mu_{B}$), and
there is no effect of the $U$ parameter, whereas the MM of the M $=$ Ti system is higher than for the isolated
molecule ($2\,\mu_{B}$), inceasing with $U$. Apparently the d-d electron interaction of Ti enhances the charge
transfer from the gold chain to the molecule when correlation increase. For V and Cr the
MM for $U=0$ is smaller than that of the isolated molecule ($3\,\mu_{B}$ for VPc, $2\,\mu_{B}$ for CrPc).
As $U$ increases from $4$ to $8$ eV, the MMs get closer to the isolated molecule value.
For Mn junctions the MM is mainly located on Mn ($4.8\,\mu_{B}$ for plain MnPc) and hardly depends on $U$. 

For the Fe case, the magnetic moment of isolated FePc happens to be close to $2\,\mu_B$, but is found to be
strongly increased in the junction due to the overlap between Au orbitals and molecular orbitals, the MM on Fe
being roughly doubled. Apparently this coupling induces a transition from a low-spin to a high-spin state, unlike
the case of, e.g., Mn which is already in a high-spin state for isolated MnPc; cf.\ \cite{Arillo}, tables I
and II. Thus Fe is an exceptional example where the coupling to the leads have a pronounced effect on the
magnetic properties. Unfortunately, it seems not to be possible to explain this surprizing fact in terms of
the standard picture of atomic d orbitals in a square planar crystal field (see also below).

For the last five columns in the table, Co $\dots$ Ag, we note, first of all, that the total MM of the respective 
isolated molecules are close to 1, 0, 1, 0, $1\,\mu_{B}$, respectively \cite{Arillo}. Thus it is apparent that the coupling
to the leads has a significant effect in all cases for MPc, while the disturbance due to the leads is rather
small for the fluorinated counterparts (considering first $U=0$). For both, MPc and fluorinated MPc, we find
a considerable correlation dependence for Co and Ni. However, due to the strong coupling to the leads for
MPc, this dependence cannot straightforwardly explained by the electronic configurations of the respective
isolated metal atoms. On the other hand, there is hardly any $U$-dependence for Cu, Zn and Ag (filled $d$
shells).

Table \ref{table1} also shows that the magnetic moment on the central metal atom in most cases
differs considerably from the total magnetic moment. The $U$-dependence of the metal magnetic
moment generally is small, except for Co and Ni.

Because of the high electronegativity of F, the MMs of the AuS-F$_{16}$MPc-SAu systems generally are
less than or equal to those of AuS-MPc-SAu. For M $=$ Cr, Cu, and Zn the MMs of the AuS-(F$_{16}$MPc)-SAu
junctions agree with those of the isolated molecules, independent of $U$. In most cases, the MMs do not
change when increasing $U$ from 4 to 8 eV, especially in the fluorinated systems. The magnetic moments
generally are slightly higher for the interacting case ($U=4$ or 8 eV) than for zero Coulomb interaction, which
is reasonable since increasing $U$ will lead to a decrease of hybridization among the d electrons.

In order to obtain some further insight into the above results, we also studied in some detail the
charge and spin density isosurfaces of the considered systems. For example, the charge isosurfaces of isolated
ScPc was presented already in \cite{Arillo} (Fig.\ 2), with the result that there is hardly any charge on
the metal. In the transport situation, this does not change; however, due to the good coupling to the Au leads,
which is mediated by the above-mentioned addition of an S atom, some charge is being transferred to the
contact region. The good coupling is also apparent in the spin isosurface. As a function of the interaction
parameter $U$, it is found that the spin density is elongated along the transport direction, and that some spin
density is shifted towards the leads, which is a general trend obtained from the SMEAGOL data.

As another example, the spin density isosurface of AuS-AgPc-SAu is shown in Fig.\ \ref{fig2}. For clarity
of presentation, we show here and in the next figure only the spin isosurface for the molecule plus the first
Au atom on both sides, i.e.\ we cut out the remaining part of the leads. Of course, the isosurfaces were
computed for the complete system. In Fig.\ \ref{fig2} the above-mentioned good coupling is clearly
visible as a relatively strong spin contributions at and near the contact. Also apparent is the small
magnetic moment on the metal; note the tiny contribution from the minority spin on Ag. Similar to ScPc,
the $U$-dependence is smooth, following the general trends mentioned above.
\begin{figure}[htb]
\includegraphics[width=0.25\textwidth,clip]{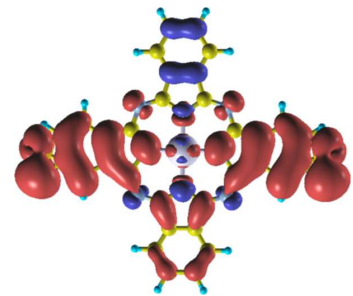}
\caption{Spin isosurface for AuS-AgPc-SAu as obtained from SIESTA (i.e., for $U=0$), taking into account
the states between $-2$ and 0 eV; cf.\ Fig.\ \ref{fig8}. Red: majority spin, blue: minority spin. Shown is
only the part containing the central molecule and the first (left and right) Au atom.}
\label{fig2}
\end{figure}

\begin{figure}[htb]
\includegraphics[width=0.22\textwidth,clip]{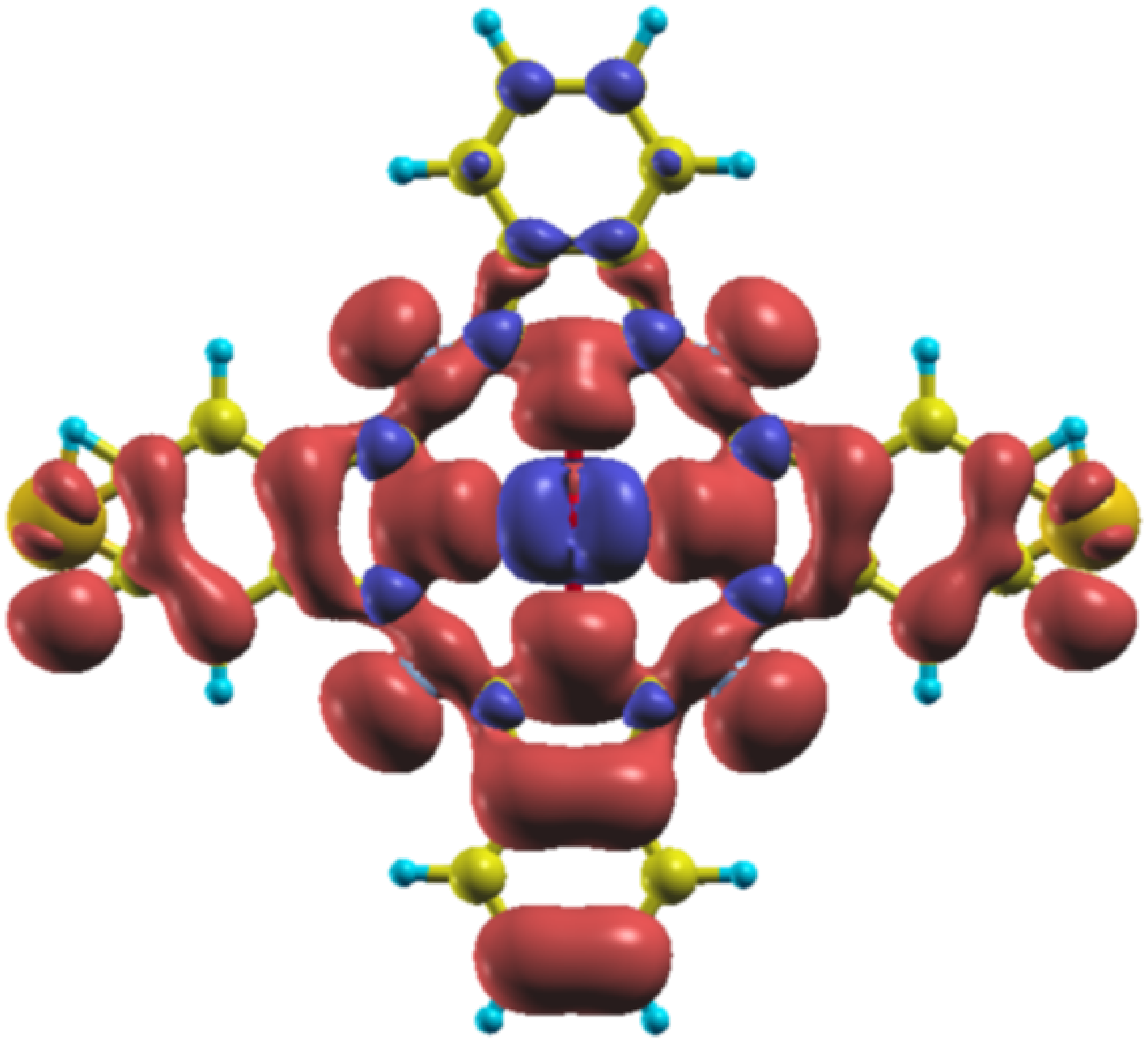}\hspace{1mm}
\includegraphics[width=0.22\textwidth,clip]{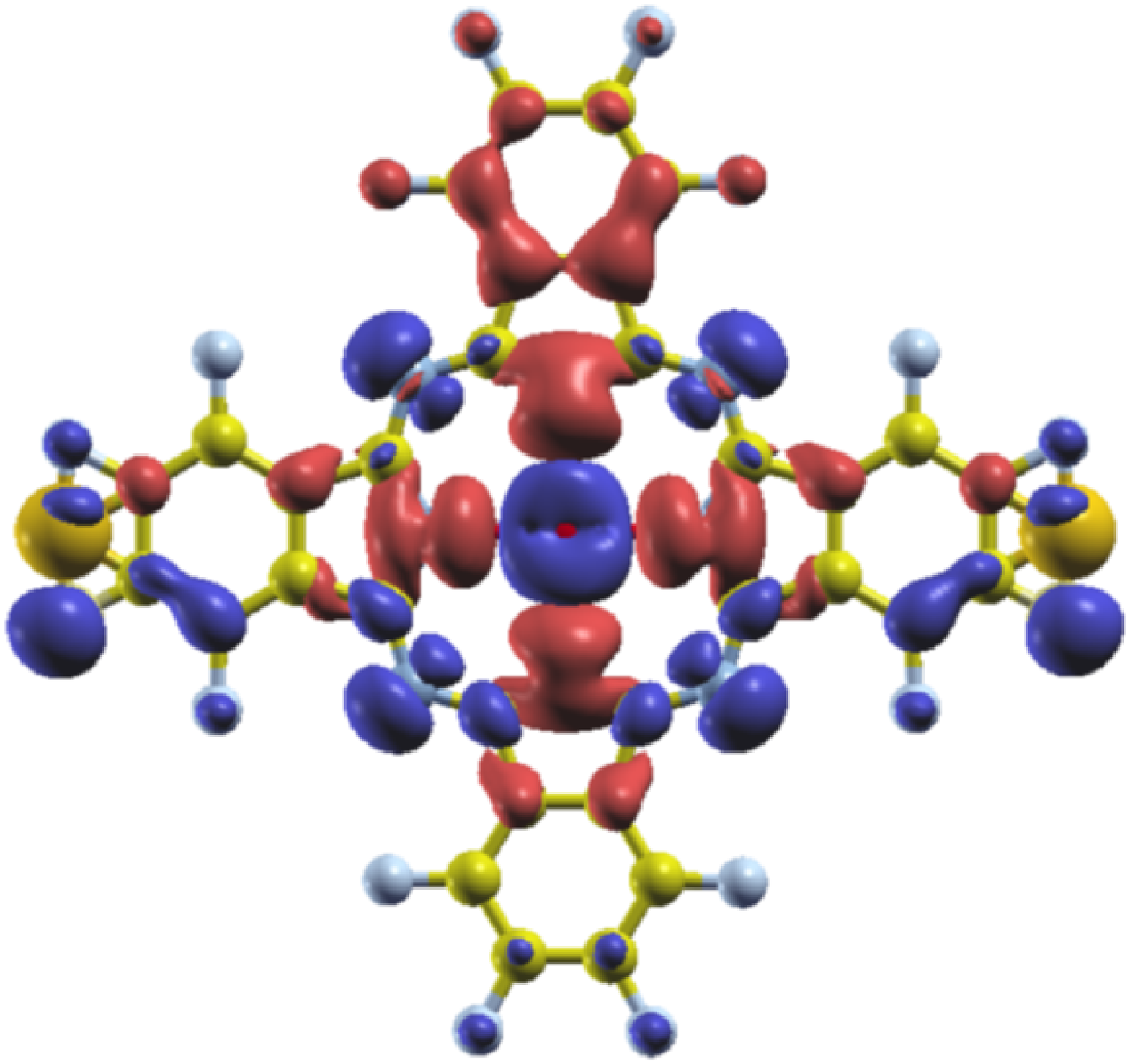}
\caption{Spin isosurface for AuS-FePc-SAu (left) and AuS-F$_{16}$FePc-SAu (right; F replacing H at the
lobes shown in grey) as obtained from SIESTA
(i.e., for $U=0$), taking into account the states between $-2$ and 0 eV; cf.\ Fig.\ \ref{fig8}.
Red: majority spin, blue: minority spin. Shown is only the part containing the central molecule and the
first (left and right) Au atom.}
\label{fig3}
\end{figure}

In Fig.\ \ref{fig3} we present the spin density isosurfaces for AuS-FePc-SAu and AuS-F$_{16}$FePc-SAu. Most
notable is the fact that the spin on Fe appears to be the same for the two cases, in agreement with the
results for the magnetic moment, cf.\ Table \ref{table1}. For FePc, one also observes large contributions
to the majority spin density from N atoms, and from C atoms which are not bonded with N. However, the C atoms which
are bonded with N as well as the central metal atom contribute to the minority spin density. There is
hardly any contribution from the H atoms. Concerning the fluorinated FePc system (right), the spin density is
clearly smaller than for FePc; cf.\ Table \ref{table1}. This reduction of the spin density can be attributed
to the reduction in the spin density of C and N atoms.

\section{Transmission properties}

In the following, we present the transmission coefficients for selected systems. Note that the d orbitals 
at the M atom have a D$_{4h}$ symmetry with a crystal field that splits the orbitals as a$_{1g}$ (d$_{z^{2}}$),
b$_{2g}$ (d$_{xy}$), e$_{g}$ (d$_{xz}$, d$_{yz}$) and b$_{1g}$ (d$_{x^{2}-y^{2}}$) states. For some systems, like
F$_{16}$ScPc and TiPc, $T(E)$ is found to be negligible: the projected density of states (PDOS) of these junctions
are mostly formed by sharp lines, which indicates a negligible coupling of molecular orbitals with the atomic
orbitals in the leads. In the following, we concentrate on a few representative cases.

\begin{figure}[htb]
\includegraphics[width=0.48\textwidth,clip]{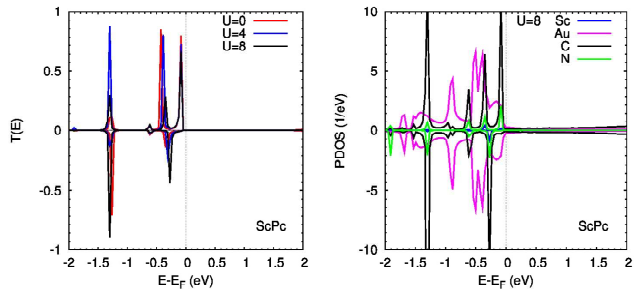}
\caption{Left: Transmission coefficient for AuS-ScPc-SAu; right: corresponding projected density of states.
Note that the Au PDOS has been scaled by a factor 0.1 for easier comparison.}
\label{fig4}
\end{figure}

{\em ScPc}.
The transmission coefficient obtained for the AuS-ScPc-SAu system is shown in Fig.\ \ref{fig4}.
The spin components are distinguished by positive and negative values.
A tiny shift of the $T(E)$ peaks is noticeable as a function of $U$, see left hand side of Fig.\ \ref{fig4}. The
PDOS for $U=8$ eV on the right hand side of Fig.\ \ref{fig4}
shows that the C and Au states dominate in the energy range $-2$ and $2$ eV, whereas N and Sc states appear
only below $E_{F}$.
The a$_{1g}$ states are part of the HOMO and the e$_{g}$ states are part of HOMO$-$1 spin
down and HOMO$-$2 spin up. The transmission peaks at $-0.2$, $-0.3$ and $-0.4$ eV are mainly due to HOMO, HOMO$-$1 and HOMO$-$2,
respectively, which are localized on Sc and Pc. 
The transmission at $-1.3$ eV is dominated by the molecular orbitals for both spins. The overlap between molecular 
orbitals and Sc orbitals is a signature that can be observed in $T(E)$ as the spatial overlap region becomes larger.
The changes in $T(E)$ when increasing the interaction parameter are very small.

\begin{figure}[htb]
\includegraphics[width=0.48\textwidth,clip]{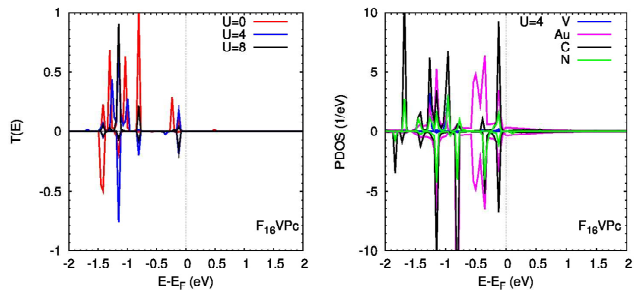}
\caption{Left: transmission coefficient for AuS-F$_{16}$VPc-SAu. Right: corresponding projected density of states}
\label{fig5}
\end{figure}

{\em Fluorinated VPc}.
Figure \ref{fig5} shows $T(E)$ and the PDOS for AuS-F$_{16}$VPc-SAu where all transmission peaks are below the Fermi energy. 
We find small variations in $T(E)$ for energies below $-0.7$ eV as the value of $U$ is changed. This is likely due to the fact that when $U$ increases, 
electronic localization is favored and the transmission becomes more localized in a smaller energy window, as shown in the left hand side of 
Fig.\ \ref{fig5}. Near the Fermi energy the direction of spin and the corresponding DOS strongly depend on $U$. The projected 
density of states for $U=4$ eV is depicted on the right hand side of Fig.\ \ref{fig5}. The degenerate e$_{g}$ states contribute 
to all molecular states along the junction and also to the transmission coefficient below the Fermi energy. At $-0.8$ and $-0.9$ eV we 
find contributions mainly from the b$_{1g}$ orbital.

We note in passing that the transmission behavior for AuS-MnPc-SAu and AuS-F$_{16}$MnPc-SAu is similar to the
case of ScPc and and fluorinated VPc. The transmission coefficient is practically independent of $U$, and very quite
small. For a different lead material (namely a SWNT), however, it has been suggested \cite{Shen} that MnPc
can be used as spin filter.

{\em FePc}.
We now consider Fe in the center of the Pc molecule. The effect of the $U$ parameter is very clear in this case. 
Some transmission coefficient peaks happen to be independent of $U$, in particular, 
those at $-1.3$ (e$_{g}$ and molecular states) for both spins, $-0.4$ eV (a$_{1g}$b$_{2g}$ and molecular 
states), and $-0.1$ eV (b$_{1g}$ and all other states of Au and the molecule) for spin up. 
From $-1$ to $-0.7$ eV, some $T(E)$ peaks appear only for $U=4$ eV, and at $-0.4$ eV for $U=8$ eV only in case
of spin down. Furthermore, at $0.9$ eV (e$_{g}$b$_{1g}$) but only for spin down, there is a single $T(E)$ peak
for $U=0$. For $U=0$ and $8$ eV, $T(E)$ has peaks at $-0.6$ eV (e$_{1g}$b$_{1g}$ and molecular states).
We find zero transmission at the Fermi energy, in contrast to the case where FePc is connected to
a SWNT \cite{Shen,Huan}.
\begin{figure}[htb]
\includegraphics[width=0.48\textwidth,clip]{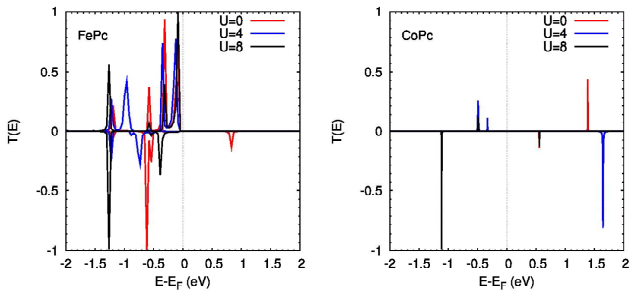}
\caption{Transmission coefficient for (left) AuS-FePc-SAu and AuS-CoPc-SAu (right)}
\label{fig6}
\end{figure}

{\em CoPc}.
For the case of Co in the center of the Pc molecule, the $T(E)$ peaks are sharp lines, and their positions vary with 
$U$. There are two peaks below and one peak above $E_F$ for every $U$ value, see right hand side 
of Fig.\ \ref{fig6}. For $U=0$, the peak in $T(E)$ at $-0.5$ eV is related to molecular as well as Au and b$_{1g}$ states.
There are two additional peaks at $0.5$ eV (molecular orbitals with b$_{1g}$) and at $1.4$ eV (molecular orbitals with
e$_{g}$b$_{1g}$a$_{1g}$).

\begin{figure}[htb]
\includegraphics[width=0.24\textwidth,clip]{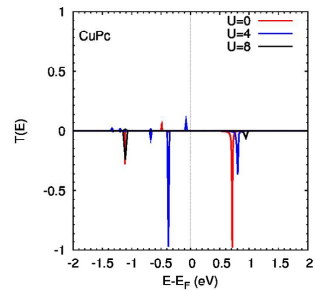}
\caption{Transmission coefficient for AuS-CuPc-SAu}
\label{fig7}
\end{figure}

{\em CuPc}.
For the AuS-CuPc-SAu junction, several transmission peaks are found below the Fermi energy, and, in particular, they are 
enhanced for $U=4$ eV, see Fig.\ \ref{fig7}. Above the Fermi energy, there is a single peak, which changes position and
intensity as function of $U$, the position shifting away from $E_F$ as $U$ increases (because of inceasing repulsion
between electrons). The energy difference between the first peak below and the first one above E$_{F}$ for $U=4$ is smaller
than the corresponding one for $U=0$ and $8$ eV.

\begin{figure}[t]
\includegraphics[width=0.48\textwidth,clip]{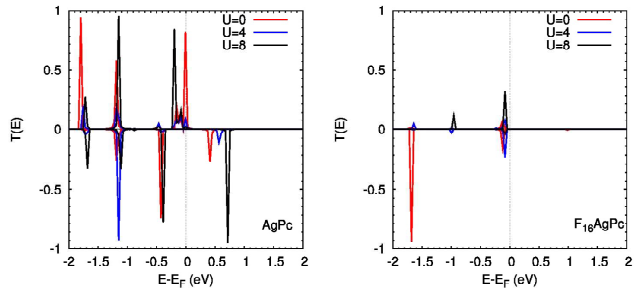}
\caption{Transmission coefficient for AuS-AgPc-SAu (left) and AuS-F$_{16}$AgPc-SAu (right)}
\label{fig8}
\end{figure}

{\em AgPc}.
Last but not least we discuss AuS-AgPc-SAu and AuS-F$_{16}$AgPc-SAu (Fig.\ \ref{fig8}). 
The $T(E)$ peaks demonstrate a good coupling between lead and molecule as shown in the left hand side of
Fig.\ \ref{fig8} (AuS-AgPc-SAu). We find peaks for spin up states at $E_F$, independent of $U$,
which indicates that this configuration has a high potential as a spintronic device. The change of
the interaction parameter does not affect the positions of the peaks below the Fermi energy. However, we
observe a strong dependence of the peak position and intensity on $U$ above the Fermi energy:
as $U$ increases, the peak moves further away from $E_{F}$, and the intensity increases almost by a factor
of two. This indicates that as electron localization increases, the transmission also increases for that
particular state. For the AuS-F$_{16}$AgPc-SAu junction, right hand side of Fig.\ \ref{fig8}, the
transmission peak near $-0.1$ eV is finite for both spin polarizations, and independent of $U$.
The transmission near the Fermi energy is dominated by molecular states with a$_{1g}$b$_{1g}$ orbitals.

For completeness we discuss the occupation of the d orbitals on the metals which were
addressed in detail in this section. First, we find only minor differences
between MPc and their fluorinated counterparts. Second, the occupation numbers of the five orbitals
add up to numbers close to the values to be expected from the periodic table:
Sc: $\approx 0.7$, decreasing by about 20 \% with increasing $U$;
V: $\approx 3.1$, almost independent of $U$;
Fe: $\approx 5.5$, decreasing by about 10 \% with increasing $U$;
Co: close to 7, independent of $U$;
Cu: $\approx 9.3$, almost independent of $U$;
Ag: $\approx 9.4$, increasing by about 1 \% with increasing $U$.
Considering the simple---from an atomic point of view---case of Sc, it is already apparent that
an interpretation in terms of the standard level picture is incomplete. This conclusion is
further supported by the case of Fe, which has a rather high magnetic moment as discussed above:
we find that all five d orbitals are occupied by close to one up electron, while the spin-down contribution
is mainly due to an e$_g$ orbital ($\approx 0.5$ for $U=4$ eV, dropping to $\approx 0.1$ when $U$
is increased to 8 eV, consistent with the increase of the metal magnetic moment from 4.2 to
$4.6\,\mu_B$, cf.\ Table \ref{table1}). For the V system, the d orbital occupations depend only
weakly on interaction: the e$_{g}$ (up) and b$_{1g}$ (up) occupations are slightly increased 
when $U$ is increased, while the others decrease by a small amount. For the Co system, we find the
up occupations to be close to one, while the occupations of the b$_{2g}$ (down), a$_{1g}$ (down),
and one of the e$_g$ (down) orbitals are considerably smaller. For Cu and Ag, all occupations
are close to one, except for the a$_{1g}$ (down) orbital ($\approx 0.6$ and $\approx 0.7$, respectively).

\section{Spin filter efficiency and conductance}

The spin filter efficiency ($SFE$) for an electronic device, defined as
\begin{equation}
\label{eq1}
SFE = \frac{T_{\uparrow}(E_{F})-T_{\downarrow}(E_{F})}{T_{\uparrow}(E_{F})+T_{\downarrow}(E_{F})} ,
\end{equation}
is the ability of a device to pass a particular spin component. Here, $T_{\uparrow}(E_{F})$ and
$T_{\downarrow}(E_{F})$ denote the transmission coefficients of spin up and spin down electrons at the
Fermi energy, respectively. Our findings are summarized in Table \ref{table2}, the positive values
corresponding to the case where ``up'' is the majority spin, while a negative sign means that
the majority spin is ``down''. We note that the magnitude and the sign of the spin filter efficiency can
vary with the interaction parameter. In some cases the majority spin in F$_{16}$MPc is opposite to the
majority spin in MPc, e.g., for Sc and Cu. Our results suggest that junctions containing Sc and Ag, as well
as FePc, ZnPc, F$_{16}$TiPc, F$_{16}$VPc and F$_{16}$CuPc could potentially be used in spintronic devices.
Note, however, that ScPc is chemically unstable \cite{Clarisse,Li}. In several of the considered molecules
we observe zero spin filter efficiency (independent of $U$), in particular, for CrPc, CoPc, NiPc, and their
fluorinated counterparts.

\begin{table*}[htb]
\caption{Spin filter efficiency for AuS-MPc-SAu and AuS-F$_{16}$MPc-SAu in $\%$}
\centering
{
\begin{tabular}{|c|c|c|c|c|c|c|c|c|c|c|c|c|}
\hline
&&Sc& Ti& V& Cr& Mn&Fe& Co& Ni& Cu& Zn& Ag\\\hline
&$U$=0
&$-$18.0 &0 &0&0&0&99.9&0&0&0&0  &99.9 \\

AuMPc&$U$=4
&$-$3.0&0 & 0&0&0&51.7&0&0&8.3&68.4  &99.9\\

&$U$=8
&$-$16.0 &0 & 0&0&0&$-$2.5&0&0&0&72.9   &97.9 \\\hline

&$U$=0
&48.8&99.9&14.2&0&0&0 &0&0&0  &0&99.5 \\

AuF$_{16}$MPc&$U$=4
&70.4&98.0&64.3&0&0&0&0 &0&$-$99.9  &0&99.2 \\
\raisebox{2ex}&$U$=8
&74.1&99.0&$-$90.6&0&0&0&0&0&$-$99.3  &0&98.6 \\\hline
\end{tabular}
}
\label{table2}
\end{table*}

However, a reasonable electronic conductance is also essential for the performance of a device,
hence we determine in addition the conductance,
\begin{equation}
\label{eq2}
G=G_{0} \, [T_{\uparrow}(E_{F})+T_{\downarrow}(E_{F})], \; G_{0}=2e^{2}/h .
\end{equation}
For the uncorrelated case, $U=0$, the junction which includes AgPc shows the highest 
conductance, close to 0.22 $G_{0}$. The next higher one is CuPc with $1.8 \times 10^{-4} \, G_{0}$. 
This value is smaller than the experimental result found in aromatic molecules by two orders of magnitude \cite{BXu,XXi},
but our result roughly agrees with another theoretical calculation \cite{Tada1}. Differences in theoretical 
results generally depends on the construction of the junctions, as well as the employed calculational method and 
the details of the (weak) interaction between molecule and leads. From a practical point of view, the results will
also depend on the interaction between the junction and the substrate, which is not taken into account here and
in most other studies.

The conductance of ZnPc is found to be $3 \times 10^{-5} \, G_{0}$, which is the same order as previous experimental
and theoretical results for an oligo-porphyrin molecular wire by using Zn in the molecule center \cite{GSe}. We
mention in passing that for TiPc and VPc, which are also chemically unstable \cite{Li,Chen,HTada,Barlow,YLPan},
a vanishing conductance is obtained.

In F$_{16}$MPc, the conductances in most cases are
less than for the corresponding MPc. The highest conductance is found for 
F$_{16}$AgPc, namely 0.002 $G_{0}$, the next highest value being $1.5 \times 10^{-4} \, G_{0}$ for F$_{16}$FePc.
In addition, the conductance of F$_{16}$NiPc and F$_{16}$CuPc is of the same order as the conductance of F$_{16}$FePc. 

The electronic conductance is somewhat insensitive to an increase of $U$ for all systems, in particular, it remains
zero for thoses cases where we find zero conductance at $U=0$. On the other hand, when the conductance is finite at $U=0$,
we find it either to remain constant or to decrease with increasing $U$. For example, for $U=4$, the conductances for
CuPc, ZnPc and AgPc are $6 \times 10^{-8}$, $3 \times 10^{-6}$ and $1.1 \times 10^{-2} G_{0}$, respectively. When we
increase $U$ to 8 eV the conductances for the same systems drop to 0, $3 \times 10^{-8}$ and $7.3 \times 10^{-3} G_{0}$,
respectively.
 
\section{Summary}
In this paper we studied systematically the magnetic, electronic and transport properties of metal phtahlocyanines
and fluorinated metal phthalocyanines connected to Au leads, including some MPc molecules which are chemically unstable.
We employed the LDA+$U$ approach, with values of the Coulomb interaction parameter
ranging from $U=0$ to $U=8$ eV. The magnetic moments are largely determined by the hybridization between
d metal and Au states near the Fermi energy. The magnetic moments (Table \ref{table1}) to some extent vary with the
electron-electron interaction on the central metal atom, $U$, which also for some systems considerably
modifies the spin filter efficiency (Table \ref{table2}). Considering Eq.\ (\ref{eq1}), it is
apparent that the spin filter efficiency is a very sensitive quantity whenever the transmission
coefficients for both channels are very small.

In particular, the magnetic moments are found to increase with increasing correlation, or are roughly
independent of $U$. For the transport properties, the situation is less clear: it appears that for some systems
the spin filter efficiency hardly depends on $U$, while others are quite sensitive to correlation effects. There
can be a remarkable difference between MPc and its fluorinated counterpart. In detail, the results
can be explained by the respective electronic structure near the Fermi surface, which, however, is not
only determined by the orbitals of the central metal atom but also by the coupling to the leads. In order
to finally clarify the role of electronic correlation in the series of junctions considered, additional
experimental efforts will be needed. Overall our theoretical results compare well with
other theoretical and experimental results for the electronic conductance where these are available.

The structures with Ag in the center of the junction are a notable exception: they show both
a good spin filter efficiency {\em and} a reasonable electronic conductance since the contribution of Ag states at
the Fermi energy is large. Hence AgPc and F$_{16}$AgPc junctions can potentially be used in spintronic devices.

\acknowledgments{We thank A.~A.~Maarouf, I.~Rungger, and C.~Schuster for fruitful discussions, and acknowledge
financial support by the 
Deutsche Forschungsgemeinschaft (through TRR 80). A.~H.~Romero  acknowledges the support by the Marie
Curie Actions from the European Union in the international incoming
fellowships (grant PIIFR-GA-2011-911070), and the Donors of the American Chemical Society Petroleum
Research Fund under contract 54075-ND10 for partial support of this research.}


\begin{thebibliography}{10}
\bibitem{note} Surprisingly, there was a convergence problem with the 300 Ry cutoff for F$_{16}$TiPc. In this case, we managed with a cutoff of 250 Ry.
\end{thebibliography}
\end{document}